\documentclass[prb,twocolumn,aps,superscriptaddress,showpacs,floatfix]{revtex4}
\usepackage{amsmath}
\usepackage{graphicx}
\usepackage{color}
\begin{document}

\title{Majorana fermions in pinned vortices}

\author{A.L. Rakhmanov}
\affiliation{Advanced Science Institute, The Institute of Physical
and Chemical Research (RIKEN), Saitama, 351-0198, Japan}
\affiliation{Institute for Theoretical and Applied Electrodynamics
Russian Acad. Sci., 125412 Moscow, Russia}

\author{A.V. Rozhkov}
\affiliation{Advanced Science Institute, The Institute of Physical
and Chemical Research (RIKEN), Saitama, 351-0198, Japan}
\affiliation{Institute for Theoretical and Applied Electrodynamics
Russian Acad. Sci., 125412 Moscow, Russia}

\author{Franco Nori}
\affiliation{Advanced Science Institute, The Institute of Physical
and Chemical Research (RIKEN), Saitama, 351-0198, Japan}
\affiliation{Department of Physics, University of Michigan, Ann
Arbor, MI 48109-1040, USA}

\begin{abstract}
Exploiting the peculiar properties of proximity-induced superconductivity
on the surface of a topological insulator, we propose a device which allows
the creation of a Majorana fermion inside the core of a pinned Abrikosov
vortex. The relevant Bogolyubov-de Gennes equations are studied
analytically. We demonstrate that in this system the zero-energy Majorana
fermion state is separated by a large energy gap, of the order of the
zero-temperature superconducting gap $\Delta$, from a band of
single-particle non-topological excitations. In other words, the Majorana
fermion remains robust against thermal fluctuations, as long as the
temperature remains substantially lower than the critical superconducting
temperature. Experimentally, the Majorana state may be detected by
measuring the tunneling differential conductance at the center of the
Abrikosov vortex. In such an experiment, the Majorana state manifests
itself as a zero-bias anomaly separated by a gap, of the order of $\Delta$,
from the contributions of the nontopological excitations.
\end{abstract}

\maketitle

\section{Introduction}

A Majorana fermion is an unconventional quantum state with non-Abelian
statistics. Until recently, the condensed matter community viewed it only
as a mathematical tool designed to help solving some specific many-body
problems, arising, for example, in the areas of the two-channel Kondo model
\cite{Kivelson_Phys.Rev.B_1992,Rozhkov_Int.J.Mod.Phys.B_1998,
Zarand_Phys.Rev.B_2000} and the quantum magnetism
\cite{Shastry_Phys.Rev.B_1997}.

However, the study of topological quantum computing
\cite{Nayak_Rev.Mod.Phys._2008}
initiated a search for experiments where this state can be directly
observed and manipulated. Several proposals have been put forward. They
rely of a diverse set of systems: liquid helium
\cite{Chung_Phys.Rev.Lett._2009,Volovik_JETPLett._2009},
topological insulators (TI)
\cite{Benjamin_Phys.Rev.B_2010},
superconducting heterostructures
\cite{alicea_device},
$p_x+ip_y$-wave superconductors
\cite{PhysRevLett.101.267002,Kraus_Phys.Rev.B_2009},
non-centrosymmetric superconductors
\cite{Fujimoto_Phys.Rev.B_2008,Sato_Phys.Rev.B_2009},
proximity-induced superconductivity on the surface of TI
\cite{fu_kane_device,sau_robustness};
Refs.~\onlinecite{sato}
studied non-abelian topological orders and Majorana fermions in s-wave
superfluids of ultracold fermionic atoms and also spin-singlet
superconductors with the spin-orbit interactions.

In this paper we discuss a Majorana state localized at the core of an
Abrikosov vortex residing in a two-dimensional (2D) superconductor.
Clearly, not every superconductor has such a state inside its Abrikosov
vortices: ordinary $s$-wave superconductors, for example, do not. Yet,
in the theoretical literature several superconducting systems are discussed
where a vortex can trap a Majorana state
\cite{Ivanov_Phys.Rev.Lett._2001,PhysRevLett.101.267002,
Kraus_Phys.Rev.B_2009}.
However, these proposals have one serious drawback: in addition to
the Majorana fermion, inside the normal core of the vortex, numerous
non-topological Caroli-de Gennes-Matricon (CdGM) states are localized as
well
\cite{CdGM}.
These states are separated from the zero energy by a minigap $\delta$ whose
size can be estimated as:
\begin{eqnarray}
\delta \sim \frac{\Delta^2}{\varepsilon_F},
\label{minigap}
\end{eqnarray}
where $\Delta$ is the superconducting gap and
$\varepsilon_F$
is the Fermi energy. In order for a device to be a building block of a
topological quantum computer it is necessary to freeze out all
non-topological degrees of freedom, that is, the operational temperature
should be much smaller than $\delta$. Since
$\varepsilon_F \gg \Delta$,
the minigap is expected to be extremely low (of the order of
$10^{-2}-10^{-3}$~K
for usual s-wave superconductors). This means that such proposals have very
dim prospects, unless a way of increasing $\delta$ is found (however, see
Ref.~\onlinecite{akhmerov}).

A possible way to overcome this shortcoming is described in
Ref.~\onlinecite{sau_robustness}
(see also
Ref.~\onlinecite{sato}).
In this reference the idea of a ``robust" Majorana fermions is put forward:
the Majorana state is robust if the eigenenergy of the lowest
non-topological excitation is of the order of $\Delta$, that is
$
\delta \sim \Delta.
$
It is found numerically that, if conditions are right, the robust Majorana
fermion exists in a vortex residing in a proximity-induced superconductor
on the surface of a TI. This result implies that the Majorana fermion in
such a system can be created and manipulated at experimentally achievable
temperatures.

A device suitable for this task is presented in
Ref.~\onlinecite{sau_robustness}
as well. It relies on two coupled tri-junction devices in which a Josephson
vortex is inserted. A tri-junction is a meeting point of three Josephson
junctions separating three superconducting islands placed on the surface of
the TI. Altogether, the system consists of four superconducting islands and
four superconducting loops with magnetic fluxes to control the
superconducting phases on the islands. The Majorana fermion is bound to
the Josephson vortex. Varying the relative phases with the help of the
fluxes one can move the fermion from one tri-junction to another.

Here we discuss a much simpler system in
which the robust Majorana fermion may exist, as shown in
Fig.~\ref{system}.
It is related to the proposal of
Ref.~\onlinecite{sau_robustness}:
the most basic component is the vortex inserted into the superconductor
induced on the TI surface by the proximity effect. It is demonstrated below
that such a vortex can host a robust Majorana state whose presence can be
detected with the help of local tunneling experiments. Thus, our proposed
setup can provide a proof-of-principle that a robust Majorana fermion is
indeed possible and robust, as claimed. However, the simplicity comes at a
price: unlike the device of
Ref.~\onlinecite{sau_robustness},
our Majorana fermion is pinned in space.

In addition to that, our results are as follows. We investigate the
Bogolyubov-de Gennes (BdG) equations which describe an Abrikosov vortex in
our system and analytically demonstrate that, indeed, the gap, separating
the Majorana state and the lowest non-topological excitation is of the
order of $\Delta$, in agreement with numerical results
\cite{sau_robustness}.
Further, we provide simple arguments explaining why the robustness of the
Majorana fermion exists in this system: it is a consequence of the
vanishing density of states in the TI.

Our paper is organized as follows. In Section II we review the derivation
of the BdG equations on the surface of the TI for the uniform case and
derive BdG equations in the presence of an Abrikosov vortex. In Section III
we analytically obtain the zero-energy solution (Majorana state) of these
equations. In Section IV, excited states of the model and the robustness of
the Majorana fermion are analyzed. In Section V we discuss the results
obtained here. More technical points are relegated to two appendices.

\section{Bogolyubov-de Gennes equations}
\label{sect::majorana}

\subsection{General formalism}

\begin{figure}[t]
\centering
\leavevmode
\includegraphics[width=8.5 cm]{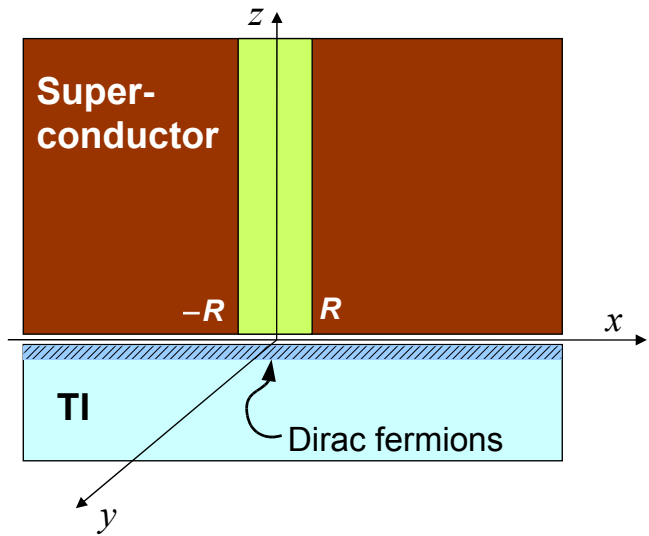}
\caption[]
{(Color online) Schematic side view of the system studied in this work. On
the surface of the topological insulator (TI) there exists a 2D band of
surface states which may be described by massless Dirac equation. To induce
superconductivity in this 2D metal, a slab of superconducting material
(SC) is placed on top of the topological insulator's surface. A cylindrical
cavity of radius $R$ is carved in the superconductor. It serves as a
pinning site for the Abrikosov vortex, which is introduced in order to
create the Majorana state. The main purpose of the cavity is to remove the
unwanted single-particle excitations in the normal core of the vortex.
Therefore, a robust Majorana fermion can be localized inside the vortex
core created on the surface of the topological insulator by the proximity
effect.
}\label{system}
\end{figure}

The system under investigation is schematically presented in
Fig.~\ref{system}.
It consists of a TI sample on which a slab of s-wave superconducting (SC)
material is placed. On the surface of the TI a 2D band of electron states
exists. This band is described by the massless Dirac equation (Weyl-Dirac
equation). The proximity to the superconductor induces a finite gap
$\Delta_{\rm TI}$
in the Dirac band.

When such a system is placed into a transverse magnetic field of sufficient
magnitude an Abrikosov vortex enters it. This vortex is accompanied by a
``pancake" vortex inside the 2D Dirac band of TI. Such a pancake vortex can
host a Majorana state
\cite{Cheng_Phys.Rev.B_2010}.
Since the core of the Abrikosov vortex contains a large amount of CdGM
states separated by small energy gap $\delta$, the Majorana state is
not robust. To counteract this disadvantage a cylindrical channel of radius
$R > \xi$
is carved in the superconductor (see
Fig.~\ref{system}).
The cavity removes the CdGM states from the vortex core. It also acts as a
pinning center for the vortex.

When the CdGM states are absent, the remaining low-lying states may be
present only inside the ``pancake" vortex core in the TI. To find them
we derive the effective BdG equations for the TI degrees of freedom. To
this end we consider the
Hamiltonian~\cite{sau_robustness}
describing the proximity effect at the TI-SC interface
\begin{eqnarray}
H=H_{\textrm{TI}}+H_{\textrm{SC}}+\widehat{\cal{T}}+\widehat{\cal{T}}^\dag,
\label{H_micro}
\end{eqnarray}
where
$H_{\textrm{TI}}$,
$H_{\textrm{SC}}$
are the Hamiltonians for the TI surface and the BCS s-wave
superconductor, and
$\widehat {\cal{T}}$ ($\widehat {\cal{T}}^\dag$)
accounts for the tunneling from the TI surface to the SC (from the SC to
the TI surface). The excitation spectrum of the model is described by the
equation
\begin{eqnarray}
\label{H}
H_{\rm TI}
\Psi_{\rm TI}
+
\widehat{\cal T}^\dag \Psi_{\rm SC}
=
\omega \Psi_{\rm TI}
\\
H_{\rm SC}
\Psi_{\rm SC}
+
\widehat {\cal T} \Psi_{\rm TI}
=
\omega \Psi_{\rm SC},
\end{eqnarray}
where
$H_{\rm TI,SC}$
and
$\widehat {\cal T}$
are written as a
$4\times 4$
matrix in the Nambu basis
\cite{sau_robustness}
\begin{eqnarray}
H_{\textrm{TI}}
&=&
\left[
	iv\mathbf{\sigma}\cdot\mathbf{\nabla}_\mathbf{r}-U ({\bf r})
\right]\tau_z\,,
\label{dirac}
\\
H_{\textrm{SC}}
&=&
-
\left(
	\frac{\mathbf{\nabla}^2_{\bf R}}{2m}+\varepsilon_F
\right)\tau_z
+
\Delta'({\bf R})\tau_x
+
\Delta''({\bf R})\tau_y,
\label{HH}
\\
\widehat {\cal T}
&=&
\tau_z {\cal T} ({\bf r-r}').
\label{tunneling}
\end{eqnarray}
In these equations
$\hbar=1$;
${\bf R} = (x, y, z)$
is the 3D coordinate inside the SC;
$\mathbf{r}=(x,y)$
is the 2D coordinate on the surface of the TI;
$\mathbf{\sigma}$
and
$\mathbf{\tau}$
are the Pauli matrices acting in the spin and charge spaces, respectively.
The parameter $v$ is the effective electron velocity at the TI surface;
$\varepsilon_F$
is the Fermi energy in the SC. The Fermi level
$U({\bf r})$
in the topological insulator may be inhomogeneous: it depends on the
external potential and the tunneling operator (see
Appendix~\ref{app::proximity}).
The tunneling kernel
${\cal T} ({\bf r-r}')$
is independent of spin and charge indices.

The wave functions
$\Psi_{\rm TI, SC}$
are the 4-component spinors:
\begin{eqnarray}
\label{nambu_spinor}
\Psi_{\rm TI, SC}
=
\left[
	u_\uparrow,
	u_\downarrow,
	v_\downarrow,
	-v_\uparrow
\right]^T.
\end{eqnarray}
The spinor
$\Psi_{\rm TI}
=
\Psi_{\rm TI} ({\bf r})$
corresponds to the surface state and depends on $x$ and $y$ only.
The spinor
$\Psi_{\rm SC}
=
\Psi_{\rm SC} ({\bf R})$
describes electrons in the superconductor bulk. It vanishes for
$z \leq 0$.

The complex order parameter in the SC is
$
\Delta = \Delta' + i \Delta'',
$
where both
$ \Delta'$
and
$ \Delta''$
are real. The superconductor is characterized by the correlation length
\begin{eqnarray}
\xi
&=&
\frac{v_{\rm F}}{|\Delta|}.
\end{eqnarray}
Starting from Eq.~(\ref{dirac}) and Eq.~(\ref{HH}), the effective BdG
equations for the TI states can be derived:
\cite{sau_robustness}
\begin{eqnarray}
\left[
	H_{\rm TI}
	-
	\widehat {\cal T}^\dag
	(H_{\rm SC} - \omega)^{-1}
	\widehat {\cal T}
\right]
\Psi_{\rm TI}
=
\omega
\Psi_{\rm TI}.
\label{BdG_symbolic}
\end{eqnarray}
Here the expression
$(H_{\rm SC} - \omega)^{-1}$
is the single-electron Green's function for the superconductor. For
details, see
Appendix~\ref{app::proximity}.

\subsection{Uniform system}
\label{subs::uniform_system}

The symbolic
Eq.~(\ref{BdG_symbolic})
is very general: it is valid for arbitrary
$\Delta ({\bf R})$
and
$\widehat {\cal T}$,
as long as the Green's function remains well-defined. Of course, finding
the Green's function for a non-uniform $\Delta$ may be practically
challenging. Fortunately, in the regime of interest the required Green's
function can be constructed from the knowledge of the Green's function for
a homogeneous system. Thus, as a first step, let us study the situation
when $\Delta$ and
$\widehat {\cal T}$
are uniform over the whole interface, the cavity is absent, and
$|\omega| < |\Delta|$.
The resultant BdG equation now reads
\begin{eqnarray}
\label{BdG1}
\left(H_{\textrm{eff}}-\omega\right)\Psi_{\textrm{TI}}=0,
\end{eqnarray}
where the effective Hamiltonian and its parameters are defined as
\begin{eqnarray}
\label{Hef}
H_{\textrm{eff}}
=
\left[
	i\tilde{v}(\omega)
	(\mathbf{\sigma}\cdot\mathbf{\nabla}_{\bf r})
	-
	\tilde{U}(\omega)
\right] \tau_z
\\
\nonumber
+
\tilde{\Delta}'(\omega)\tau_x
+
\tilde{\Delta}''(\omega)\tau_y,
\\\label{vef}
\tilde{v}(\omega)
=
\frac{
	v\sqrt{|\Delta|^2-\omega^2}
     }
     {
	\sqrt{|\Delta|^2-\omega^2}+\lambda
     },
\\\label{Uef}
\tilde{U}(\omega)
=
\frac{
	(U + \delta U) \sqrt{|\Delta|^2-\omega^2}
     }
     {
	\sqrt{|\Delta|^2-\omega^2}+\lambda
     },
\\
\label{Delta_eff}
\tilde{\Delta}(\omega)
=
\tilde{\Delta}'(\omega)
+
i\tilde{\Delta}''(\omega)
=
\frac{
	\Delta\lambda
     }
     {
	\sqrt{|\Delta|^2-\omega^2}+\lambda
     }.
\end{eqnarray}
The quantity $\lambda$ has the dimension of energy. It characterizes the
transparency of the barrier separating the SC and TI,
and can be measured in a tunneling experiment performed at
$T > T_c$.
For our purposes we need a sufficiently thick insulating layer between the
superconductor and the TI to guarantee the low transparency of the barrier
$(\lambda \ll \varepsilon_{\rm F})$.
\cite{sau_robustness}

The contact with the superconductor shifts the bare Fermi energy $U$ by the
amount
$\delta U = O( \lambda)$.
The details of the derivation can be found in
Appendix~\ref{app::proximity}.

For uniform $\lambda$ and $\Delta$ we can choose $\Delta$ to be real.
Eq.~(\ref{BdG1})
has no solutions with
$|\omega| < \Delta_{\rm TI}$,
where the proximity-induced gap
$\Delta_{\rm TI} < \Delta$
satisfies the equation
$\tilde \Delta(\Delta_{\rm TI}) = \Delta_{\rm TI}$,
or, equivalently,
\begin{eqnarray}
\label{gap_TI}
\frac{\lambda}{\Delta}
=
\frac{\Delta_{\rm TI}}{\Delta}
\sqrt{
	\frac{\Delta + \Delta_{\rm TI}}{\Delta - \Delta_{\rm TI}}
     }.
\end{eqnarray}
The gap in the TI is a monotonous function of $\lambda$:
$\Delta_{\rm TI}\approx\lambda$
at small
$\lambda \ll \Delta$
and approaches $\Delta$ from below at large $\lambda\gg\Delta$.

\subsection{The system with the pinned vortex}
\label{sub::deriv_abrikosov}

The system schematically drawn in
Fig.~\ref{system},
however, is not uniform. Due to the cavity and the vortex, both $\Delta$
and $\lambda$ acquire some coordinate dependence. Obviously,
$\lambda (r)$
vanishes for
$r < R$,
where
$r = \sqrt{x^2 + y^2}$
is distance to the axis of the cavity. In addition, when the vortex is
introduced~\cite{dG},
$\Delta({\bf R})$
can be written as
\begin{eqnarray}
\label{order_param}
\Delta ({\bf R})
=
|\Delta (r)|e^{i\theta},
\end{eqnarray}
where $\theta$ is the polar angle in the
$(x, y)$-plane,
and
$|\Delta (r)|$
is an increasing function of $r$, which approaches the bulk value
$|\Delta|$
when
$r \rightarrow \infty$.
It is finite at
$r = R+0$.
In such a case, strictly speaking, one has to re-calculate the
superconducting Green's function for a spatially varying
$\Delta (r,\theta)$.
This might be particularly difficult for
$r \lesssim \xi$,
where the phase $\theta$ varies quickly on the distances of the order of
$\xi$.

However, one can avoid the latter complication if
\begin{eqnarray}
\label{R_cond}
R \gg \xi.
\end{eqnarray}
In this limit our formalism can be easily adopted to account for the vortex
presence. Ignoring the detailed behavior of
$|\Delta(r)|$
when
$r \approx R$,
we assume that
\begin{eqnarray}
\label{heaviside}
|\Delta(r)| = |\Delta| \vartheta ( r - R ),
\end{eqnarray}
where
$\vartheta (r)$
is the Heaviside step-function.

When
Eq.~(\ref{R_cond})
holds true, the order parameter phase $\theta$ varies slowly on distances
of the order of $\xi$. Thus, it is permissible to insert the non-uniform
$\Delta (r, \theta)$,
Eq.~(\ref{order_param}),
directly into
Eqs.~(\ref{Hef})-(\ref{Delta_eff}).
Since
$|\Delta|$
is $r$-dependent, therefore,
$\tilde v$,
$|\tilde \Delta|$,
and
$\tilde U$
are non-uniform. Further, we assume that our treatment remains valid, at
least qualitatively, in the case
$R\gtrsim \xi$.

In our formalism the TI area beneath the cavity
($r < R$)
is non-superconducting. It may be viewed as the normal core of ``the
pancake" vortex. Outside the core
(for $r > R$),
the absolute value of the order parameter equals to its equilibrium value
$|\Delta|$.
This approximation is very natural in the case $R>\xi$.

To calculate the eigenenergies of the Hamiltonian
Eq.~\eqref{Hef}
with a vortex, it is standard to exploit the cylindrical symmetry of the
problem to separate the variables (for details see
Appendix~\ref{app::radial}).
If we define the spinor $\Phi$ as
\begin{eqnarray}
\Psi_{\textrm{TI}}
=
\exp{\left[i\theta(\tau_z+\sigma_z)/2 + i\mu \theta \right]} \Phi^\mu(r),
\\
\Phi^\mu=(f_1^\mu,f_2^\mu,f_3^\mu,-f_4^\mu)^T,
\end{eqnarray}
then, its four components satisfy
\begin{eqnarray}
\label{f1}
\nonumber
i\tilde{v}
\left(
	\frac{d}{dr}
	+\frac{\mu+1}{r}
\right)f_2^\mu
+
|\tilde{\Delta}|f_3^\mu
-
(\omega+\tilde{U})f_1^\mu
&=& 0,
\\ \nonumber
i\tilde{v}\left(
		\frac{d}{dr}-\frac{\mu}{r}
	 \right)f_1^\mu
	-
	|\tilde{\Delta}|f_4^\mu
	-
	(\omega+\tilde{U})f_2^\mu
&=& 0,
\\
i\tilde{v}\left(
		\frac{d}{dr}
		+
		\frac{\mu}{r}
	 \right)f_4^\mu
	+
	|\tilde{\Delta}|f_1^\mu
	-
	(\omega-\tilde{U})f_3^\mu
&=& 0,
\\
\nonumber
i\tilde{v}\left(
		\frac{d}{dr}-\frac{\mu-1}{r}
	 \right)f_3^\mu
	-
	|\tilde{\Delta}|f_2^\mu
	-
	(\omega-\tilde{U})f_4^\mu
&=& 0.
\end{eqnarray}
This is the most general system of equations describing the sub-gap
states near the cavity. Below we study the spectral properties of this
system.

\section{Zero-energy Majorana fermion solution}

In this section we will demonstrate that the Hamiltonian $H$ has a
zero-energy Majorana fermion solution for arbitrary $R$. First, we will
demonstrate this for the case of large $R$, when one can map $H$ on
$H_{\rm eff}$.
Afterward this result will be generalized for any
$R > 0$.
This section contains some known results (see, e.g, 
Refs.~\onlinecite{Jackiw_Nuclear,Cheng_Phys.Rev.B_2010,bergman,ghaemi}),
but these are included here to make the derivation more complete and
self-contained.

\subsection{Majorana state for large $R$}

The system of equations
(\ref{f1})
can be solved exactly when
$\omega = 0$,
$\mu = 0$,
and
$\delta U (r) = 0$.
The latter requirement is satisfied only in the presence of the external
gate potential which compensates for the Fermi level shift induced by the
coupling to the superconductor (see
Appendix~\ref{app::proximity}).
In subsection~\ref{finite_R_U}
it is demonstrated that a non-zero
$\delta U (r)$
does not destroy the
$\omega = 0$
solution.

To find the desired solution we define new functions
\begin{eqnarray}
\label{XY}
\nonumber
 X_1^\mu=if_1^\mu+f_4^\mu\,,\qquad X_2^\mu=if_1^\mu-f_4^\mu\,, \\
 Y_1^\mu=if_2^\mu+f_3^\mu\,, \qquad  Y_2^\mu=if_2^\mu-f_3^\mu\,.
\end{eqnarray}
For these functions the system of
Eqs.~\eqref{f1}
splits into two systems of two equations each:
\begin{eqnarray}
\label{XY1}
\tilde{v}\frac{dX_2^0}{dr}+|\tilde{\Delta}|X_2^0=i\tilde{U}Y_1^0\,, \\
\label{XY2}
\tilde{v}\frac{dY_1^0}{dr}+|\tilde{\Delta}|Y_1^0+
	\frac{\tilde v}{r}Y_1^0=i\tilde{U}X_2^0\,,
\end{eqnarray}
and
\begin{eqnarray}
\label{XY3}
\tilde{v}\frac{dX_1^0}{dr}-|\tilde{\Delta}|X_1^0=i\tilde{U}Y_2^0\,,\\
\label{XY4}
\tilde{v}\frac{dY_2^0}{dr}-|\tilde{\Delta}|Y_2^0+
	\frac{\tilde v}{r}Y_2^0=i\tilde{U}X_1^0\,.
\end{eqnarray}

An elementary analysis reveals that the system of
Eqs.~(\ref{XY3})-(\ref{XY4})
has no non-zero solution decaying at
$r\rightarrow\infty$.
Thus,
$X_1^0=Y_2^0=0$.
The non-zero solution of the system of
Eqs.~(\ref{XY1})-(\ref{XY2})
can be written explicitly. Keeping in mind that
$\tilde U(r)/\tilde v(r) = U/v$
is a constant independent of $r$ [see
Eqs.~(\ref{vef}) and~(\ref{Uef})],
one derives:
\begin{equation}
\label{maj}
 \left(\!
  \begin{array}{c}
    X_2^0 \\
    Y_1^0 \\
  \end{array}
  \!\right)
  \!=2C\!\left(\!
      \begin{array}{c}
        iJ_0(Ur/v) \\
        J_1(Ur/v) \\
      \end{array}
      \!\right)
\exp{\left(-\int_0^r \frac{
				dr' |\tilde \Delta (r')|
			  }
			  {
				\tilde v (r')
			  }
    \right)},
\end{equation}
where $C$ is a normalizing coefficient, and $J_0$, $J_1$ are
Bessel functions. Thus, the exact solution for
$\omega=0$
is
\begin{equation}
\label{majTI}
\!\!\Psi_{\textrm{M}} ({\bf r})\!=\!C
\left(\!\!
  \begin{array}{c}
   \exp (i\theta) J_0(Ur/v) \\
   -i J_1(Ur/v) \\
   J_1(Ur/v) \\
    i \exp (- i \theta) J_0(Ur/v) \\
  \end{array}
\!\!\right)\!
\exp{\left(-\int_0^r \frac{
				dr' |\tilde \Delta (r')|
			  }
			  {
				\tilde v (r')
			  }
    \right)}.
\end{equation}
If the ratio
$|\tilde \Delta(r)|/\tilde v(r) = \lambda / v$
is a constant independent of $r$
[see
Eqs.~(\ref{vef}) and~(\ref{Delta_eff})]
then the spinor
$\Psi_{\textrm{M}}$
decays for distances larger than
$v/ \lambda$,
which may be viewed as a characteristic localization length of the
zero-energy state.

To prove that the eigenfunction given by
Eq.~(\ref{majTI})
corresponds to the Majorana state, consider the following fermion operator
\begin{eqnarray}
\hat{\Psi}^\dag_{\rm M}
=
\int d^2 {\bf r}
\left[
	u^{\rm M}_\uparrow ({\bf r})
	\psi^\dag_\uparrow ({\bf r})
	+
	u^{\rm M}_\downarrow ({\bf r})
	\psi^\dag_\downarrow ({\bf r})
\right.
\\
\nonumber
\left.
	+
	v^{\rm M}_\uparrow ({\bf r})
	\psi^{\vphantom{\dagger}}_\uparrow ({\bf r})
	+
	v^{\rm M}_\downarrow ({\bf r})
	\psi^{\vphantom{\dagger}}_\downarrow ({\bf r})
\right],
\end{eqnarray}
where
$\psi_\sigma^\dag ({\bf r})$
is the creation operator for an electron with spin
$\sigma$
located at point
${\bf r}$.
The functions
$u^{\rm M}_\sigma ({\bf r})$,
$v^{\rm M}_\sigma ({\bf r})$
are components of the spinor
${\Psi}_{\rm M} ({\bf r})$.
The operator
$\hat{\Psi}^\dag_{\rm M}$
creates a fermion in the state corresponding to
${\Psi}_{\rm M} ({\bf r})$.
It is easy to demonstrate, by direct calculation, that
$
\hat{\Psi}^\dag_{\rm M}
=
i\hat{\Psi}^{\vphantom{\dagger}}_{\rm M}.
$
Therefore,
$\Psi_{\rm M}$
corresponds to the Majorana fermion.

\subsection{Majorana state for arbitrary $R$ and $U$}
\label{finite_R_U}

We demonstrated above that for large $R$ our system can be mapped to 2D
Dirac electrons. In the latter model, when the vortex is present,
the zero-energy solution is found
\cite{Jackiw_Nuclear}.

Unfortunately, if $R$ is small, this mapping is inapplicable. What happens
to the Majorana state when the cavity is small? Below we will prove that
our Hamiltonian has a zero-energy eigenstate for any
$R \geq 0$.

We start our reasoning with the observation that
$H$
satisfies the following charge-conjugation relation:
\begin{eqnarray}
H = - \tau_y \sigma_y H^* \tau_y \sigma_y.
\label{charge_conj}
\end{eqnarray}
Thus, for every eigenstate
$\Psi$
of $H$ with a non-zero eigenenergy
$\omega \ne 0$,
an eigenstate
$\tau_y \sigma_y \Psi^*$
with eigenenergy $-\omega$ is present. A spinor with positive eigenenergy
corresponds to the creation of a quasiparticle, while the charge-conjugated
spinor corresponds to the destruction of this quasiparticle.

Further, it is demonstrated here that for large $R$ the Hamiltonian has the
zero-energy solution
$\Psi_{\rm M}$.
This eigenstate is special for it remains unchanged after a
charge-conjugation transformation. This means that the number of
eigenenergies lying inside of the even-energy interval
$(|\Delta|, -|\Delta|)$
is odd (i.e., all the non-zero eigenstates are paired, while the Majorana
state is unpaired). If we start decreasing $R$ this property endures: due
to symmetry [see
Eq.~(\ref{charge_conj})]
the eigenstates can enter or leave our energy interval only in pairs. Thus,
for any $R$ the Hamiltonian $H$ has an unpaired
$\omega = 0$
eigenstate invariant under charge conjugation. However, it is necessary to
remember that, when
$R<\xi$,
the Majorana state cannot be robust due to the CdGM states in the Abrikosov
vortex core.

Instead of $R$, one can vary
$\delta U$.
The above reasoning can be modified to prove that the deviation of
$\delta U(r)$
from the
$\delta U(r)=0$
value does not destroy the Majorana state.

\section{Excited states in the vortex core}

In addition to the
$\omega = 0$
state, it is possible to have
$0 < |\omega| < |\Delta_{\rm TI}|$
states localized at the vortex core.

\subsection{Analytical calculations}

To find the eigenfunctions for these states, it is necessary to solve
the system~(\ref{f1})
for generic $\omega$ and $\mu$. The solution can be simplified
significantly for
$\tilde U = 0$.
We will now investigate this case. The non-zero
$\tilde{U}(r)$
may be accounted with the help of perturbation theory, at least for
$|\tilde U|\ll \tilde \Delta$.
According to Eqs.~(26) and~(27) of
Ref.~\onlinecite{sau_robustness},
the case
$|\tilde U|> \tilde \Delta$
is not favorable for the robustness of the Majorana state and will not be
studied here.

When
$\tilde{U}=0$
and
$r < R$,
the system~(\ref{f1})
decouples into two sets of equations:
\begin{eqnarray}
\label{eq1}
i{v}\left(
		\frac{d}{dr}+\frac{\mu+1}{r}
	 \right)f_2^\mu-\omega f_1^\mu
&=& 0,
\\
i{v}\left(
		\frac{d}{dr}-\frac{\mu}{r}
	 \right)f_1^\mu-\omega f_2^\mu
&=&
0,
\end{eqnarray}
and
\begin{eqnarray}
\label{eq2}
i{v}\left(
		\frac{d}{dr}+\frac{\mu}{r}
	 \right)f_4^\mu-\omega f_3^\mu
&=&
0,
\\
i{v}\left(
		\frac{d}{dr}-\frac{\mu-1}{r}
	 \right)f_3^\mu-\omega f_4^\mu
&=&
0.
\end{eqnarray}
The solution is
\begin{eqnarray}
\label{sol_eq1}
\nonumber
f_1^\mu=iA_\mu J_\mu(\omega r/{v}),
\qquad
 f_2^\mu=A_\mu J_{\mu+1}(\omega r/{v}),
\\
f_3^\mu=iB_\mu J_{\mu-1}(\omega r/{v}),
\qquad
 f_4^\mu=B_\mu J_\mu(\omega r/{v}),
\end{eqnarray}
where $A_\mu$ and $B_\mu$ are constants.

For
$r > R$ the equations are:
\begin{eqnarray}
\label{eq3}
\nonumber
Y_2^\mu
=
\frac{i\tilde{v}}{\omega}
	\left(
		\frac{d X_2^\mu}{dr}
		+
		\frac{|\tilde{\Delta}|}{\tilde{v}}X_2^\mu
		-
		\frac{\mu}{r}X_1^\mu
	\right),\qquad
\\
Y_1^\mu
=
\frac{i\tilde{v}}{\omega}
	\left(
		\frac{d X_1^\mu}{dr}
		-
		\frac{|\tilde{\Delta}|}{\tilde{v}}X_1^\mu
		-
		\frac{\mu}{r}X_2^\mu
	\right),\qquad
\\
\nonumber
\frac{d^2X_1^\mu}{dr^2}
+
\frac{1}{r}\frac{dX_1^\mu}{dr}
-
\left(
	\frac{1}{[\tilde \xi (\omega)]^2}
	+
	\frac{|\tilde{\Delta}|}{\tilde{v}r}
	-
	\frac{\mu^2}{r^2}
\right)X_1^\mu
= 0, \\
\nonumber
\frac{d^2X_2^\mu}{dr^2}
+
\frac{1}{r}\frac{dX_2^\mu}{dr}
-
\left(
	\frac{1}{[\tilde \xi (\omega)]^2}
	-
	\frac{|\tilde{\Delta}|}{\tilde{v}r}
	+
	\frac{\mu^2}{r^2}
\right)X_1^\mu
=0,
\end{eqnarray}
where the functions
$X^\mu_{1,2}$
and
$Y^\mu_{1,2}$
are defined by
Eqs.~(\ref{XY}),
and the length
$\tilde \xi (\omega)$
is given by the formula:
\begin{eqnarray}
\label{xi_omega}
\tilde \xi (\omega)
&=&
\frac{\tilde v (\omega)}{\sqrt{|\tilde \Delta (\omega)|^2 - \omega^2}}.
\end{eqnarray}
The above equations may be solved
\cite{AbrSteg}
in terms of the Whittaker functions
$W_{\alpha, \beta} (z)$:
\begin{eqnarray}
\label{whittaker}
X^\mu_{1,2}
=
\frac{C_{1,2}}{\sqrt{r}}
W_{\alpha_{1,2}, \mu}
		\left(
			2r/\tilde \xi(\omega)
		\right),
\\
\label{index_alpha}
\alpha_{1,2} = \mp \frac{
				|\tilde \Delta|
			}
			{
				2 \sqrt{ |\tilde \Delta|^2 - \omega^2 }
			}.
\end{eqnarray}
Using
Eq.~(\ref{gap_TI})
one can show that for subgap states
($|\omega| < \Delta_{\rm TI}$)
the expression under the square root in
Eqs.~(\ref{xi_omega})
and~(\ref{index_alpha})
is positive, therefore,
$\tilde \xi (\omega)$
and
$\alpha_{1,2}$
are real.
Equation~(\ref{whittaker})
implies that
$\tilde \xi (\omega)$
is the energy-dependent localization length for the subgap states.

Matching the solutions at $r=R$ we derive the equation for the subgap
eigenenergies:
\begin{eqnarray}
\label{excited1}
\left(
	\frac{W'_{\alpha_1,\mu}}
	     {\tilde \xi W_{\alpha_1,\mu}}
	+
	\frac{W'_{\alpha_2,\mu}}
	     {\tilde \xi W_{\alpha_2,\mu}}
	-
	\frac{\mu+1}{R}
	+
	\frac{\omega  J_{\mu+1}}{\tilde v J_\mu}
\right)
\\
\nonumber
\times
\left(
	\frac{W'_{\alpha_1,\mu}}
	     {\tilde \xi W_{\alpha_1,\mu}}
	+
	\frac{W'_{\alpha_2,\mu}}
	     {\tilde \xi W_{\alpha_2,\mu}}
	+
	\frac{\mu-1}{R}
	-{\omega  J_{\mu-1}\over{\tilde v J_\mu}}
\right)
\\
\nonumber
=
\left(
	\frac{W'_{\alpha_1,\mu}}
	     {\tilde \xi W_{\alpha_1,\mu}}
	-
	\frac{W'_{\alpha_2,\mu}}
	     {\tilde \xi W_{\alpha_2,\mu}}
	-
	\frac{\tilde \Delta}{ \tilde v}
\right)^2.
\end{eqnarray}
In this equation all functions
$W_{\alpha,\beta }(z)$
and
$W'_{\alpha,\beta }(z)$
must be evaluated at
$z=2R/\tilde \xi(\omega)$,
and all Bessel functions must be evaluated at
$\omega R/ v$. We analyzed Eq.~(\ref{excited1})
numerically.

\subsection{Majorana fermion robustness}

\begin{figure}[btp]
\centering
\leavevmode
\includegraphics[width=8.5 cm]{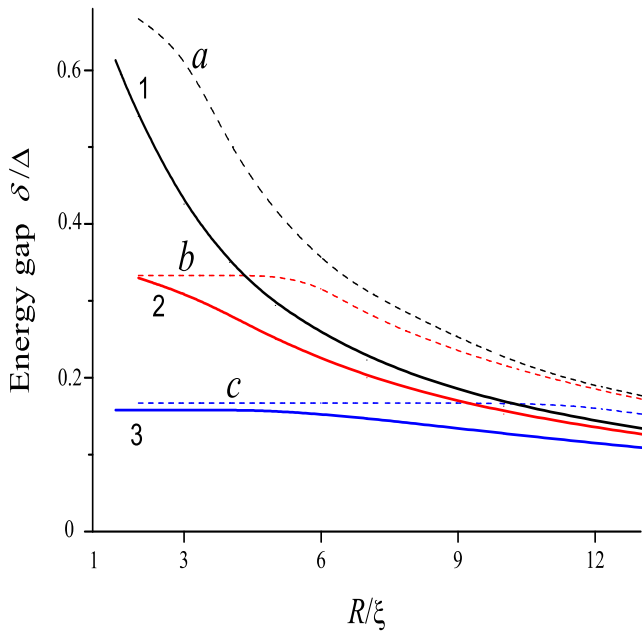}
\caption[]
{(Color online) The energy gap $\delta$ between the ground state and
excited states as a function of the cavity radius $R$, for different
transparencies $\lambda$ of the barrier. The solid curves 1, 2, and 3
correspond to the states with
$n=0,\,\mu=1$,
and the dashed curves 
\textit{a, b, c}
correspond to the states with
$n=1,\,\mu=0$.
As we go from top to bottom, the transparency $\lambda$ decreases: curves 1
and \textit{a} (black) are drawn for
$\lambda/\Delta=2$;
curves 2 and \textit{b} (red) are drawn for
$\lambda/\Delta=0.5$;
curves 3 and \textit{c} (blue) are drawn for
$\lambda/\Delta=0.1$.
}\label{excited}
\end{figure}

The excited quantum states of our system can be classified using two
quantum numbers: the radial number $n$ and the orbital number $\mu$. The
Majorana state obtained above corresponds to the
$n=0,\,\mu=0$
state. The results of the numerical solution of
Eq.~\eqref{excited1}
for different values of
($n,\,\mu$)
are shown in
Fig.~\ref{excited}.
Here the reduced gaps 
$\delta/\Delta$
between the ground state and excited states with
($n=0,\,\mu=1$)
and
($n=1,\,\mu=0$)
are plotted for different transparencies $\lambda$ as a function of the
cavity radius $R$. The solution of
Eq.~\eqref{excited1}
confirms an intuitively transparent conclusion: that the first excited
state is either the
($n=0,\,\mu=1$)
state, or the
($n=1,\,\mu=0$)
state. Note, that, to calculate $\delta$ in the case
($n=1,\,\mu=0$),
we use the asymptotics of the Whittaker functions
$W_{\alpha_{1,2},0}(z)$
valid at
$|z|\gg 1$.
This negatively affects the accuracy of our calculations at small $R$. We
believe, however, that this substitution does not distort the qualitative
features of the solution.

As it is seen from
Fig.~\ref{excited},
the state
($n=0,\,\mu=1$)
lies lower than the state 
($n=1,\,\mu=0$).
Thus, the gap between the ground state and the excited state
($n=0,\,\mu=1$)
characterizes the robustness of the Majorana fermion in our system. The gap
value increases when $\lambda$ increases. If
$\lambda \gtrsim\Delta$,
the energy gap
$\delta(R)$
practically saturates and further growth of the barrier transparency does
not significantly improve the robustness. Experimentally, the regime
$\lambda \sim \Delta$
corresponds to a barrier with low transparency
\cite{sau_robustness},
since $\lambda$ is much smaller than the Fermi energy.

At a given $\lambda$, the curve
$\delta(R)$
is a decreasing function of $R$, approaching a maximum value
\begin{eqnarray} 
\delta_{\textrm{max}}
=
\tilde{\Delta}(0)
=
\frac{
	\Delta\lambda
     }
     {
	(\Delta+\lambda)
     }
{\rm \ at \ }
R\sim\xi.
\end{eqnarray} 

The main conclusion that follows from the results shown in
Fig.~\ref{excited}
is that the energy gap between the Majorana state and the first excited
state may be of the order of $\Delta$
($\delta/\Delta =0.5$--$0.6$ 
or even higher) if
$\lambda/\Delta \gtrsim 1$--$3$
and
$R/\xi=2$--$3$.
Therefore, a suitable choice of $R$ and $\lambda$ allows one to realize the
robust Majorana state.

\subsection{Physical and intuitive explanation of the Majorana state
robustness}

Above we demonstrated that in our system a robust Majorana state exists.
Our results agree with previous numerical calculations
\cite{sau_robustness}.
However, it is desirable to have a simple non-technical physical argument
explaining these results. To this end, now consider the ``pancake" core. It
can be approximately described as a circle of radius $R$, where
the superconducting gap is zero. Let us now evaluate the number
$N^{\rm core}_{\rm TI}$
of the single-electron subgap states:
\begin{eqnarray}
N^{\rm core}_{\rm TI}
\sim
\pi R^2
\! \!
\int_0^{\Delta_{\rm TI}}
\! \! \! \!
\nu_{\rm TI} (\epsilon) \; d \epsilon
\;
\sim
\;
\frac{
	\Delta^2_{\rm TI} R^2
     }
     {
	v^2
     }.
\end{eqnarray}
Here
$\nu_{\rm TI} (\epsilon) \sim \epsilon/v^2$
is the density of states for the TI. The average energy interval between
these states is
\begin{eqnarray}
\delta_{\rm TI}
\;
\sim 
\;
\Delta_{\rm TI}/N^{\rm core}_{\rm TI}
\;
\sim
\;
v^2/(R^2 \Delta_{\rm TI}).
\end{eqnarray}
For
$R \sim \xi$
and
$\Delta_{\rm TI} \sim \Delta$,
one has
$N^{\rm core}_{\rm TI} \sim 1$
and
$\delta_{\rm TI} \sim \Delta$.
The last estimate is equivalent to the statement of the Majorana fermion
robustness.

If, instead of TI, we now consider a 2D superconductor with parabolic
dispersion, we then obtain:
\begin{eqnarray}
N_{\rm m}^{\rm core}
\;
\sim
\;
\pi R^2 \Delta \nu_{\rm m} (\varepsilon_{\rm F})
\;
\sim
\;
\left(
	\frac{\varepsilon_{\rm F}}{\Delta }
\right)
\!
\frac{ R^2 }{{ \xi}^2 }.
\label{m_state}
\end{eqnarray}
Here
$\nu_{\rm m} (\varepsilon)$
is the density of states for a 2D metal
[$\nu_{\rm m} (\varepsilon)=4m$].
When
$R \sim \xi$,
the number
$N_{\rm m}^{\rm core}$
is much greater than unity. Further, the energy difference between the CdGM
energy levels is of the order of
$( \Delta^2 / \varepsilon_{\rm F}) (\xi^2 / R^2)$.
For
$R \sim \xi$
we recover Eq.~(\ref{minigap}).

The estimate of the previous paragraphs demonstrates that the source of
robustness of the Majorana fermion in TI is the vanishing density of states
at the Dirac point. If we were to apply a bias shifting the Fermi level
away from the Dirac point, at a bias value $U$ substantially exceeding
$\tilde \Delta$,
we recover
Eq.~(\ref{minigap}). Thus, $U=0$ is a favorable condition for the Majorana state robustness.

\section{Discussion and conclusion}
\label{sect::conclusion}

We demonstrated that in the proposed system a robust Majorana state may
exist. A possible way to detect its presence is to perform the following
tunneling experiment: insert an STM tip into the cavity and measure the
electron current flowing through the TI surface. In such a setup the
Majorana state manifests itself as a zero-bias anomaly: a peak of the
differential conductance located at
$eV = 0$.
Unlike the zero-bias anomaly observed for an Abrikosov vortex in a BCS
superconductor
\cite{ZBA1,ZBA2},
in the case of the robust Majorana state, the zero-bias anomaly is accompanied
by several discrete subgap peaks which correspond to the excited states
bound in the vortex core.


We would like to compare our proposal with the system discussed in
Ref.~\onlinecite{sau_robustness}.
It is clear that in our system the Majorana fermion is pinned to the
cavity. The advantage of the setup in
Ref.~\onlinecite{sau_robustness}
is that it allows one to shift the Majorana fermion along a straight
channel over the distance
$L \sim \xi$. However, the
robustness of the Majorana state decreases when one tries to increase the
distance $L$ above $\xi$: the gap separating the Majorana fermion and the
lowest excited state can be estimated as
$\delta \sim v/L$
for
$L > \tilde \xi$.
Moreover, the complexity of their
\cite{sau_robustness}
system is an additional limitation.

To conclude, we propose a SC-TI setup in which a robust Majorana state may
be realized inside the core of the pinned vortex. The robustness was
justified with the help of both analytical and physically intuitive
arguments. An experimental detection of this state is also discussed.

\section*{ACKNOWLEDGMENT}

This work was supported in part by JSPS-RFBR Grant
No. 09-02-92114 and RFBR Grant No. 09-02-00248.
F.N. acknowledges partial support
from the National Security Agency (NSA), Laboratory Physical
Sciences (LPS), Army Research Office (ARO), DARPA, Air Force
Office of Scientific Research (AFOSR), and National Science
Foundation (NSF) grant No.~0726909, Grant-in-Aid for Scientific
Research (S), MEXT Kakenhi on Quantum Cybernetics, and Funding
Program for Innovative R\&D on S\&T (FIRST).

\appendix
\section{Derivation of the effective Hamiltonian for proximity-induced
superconductivity}
\label{app::proximity}

In this Appendix we briefly present a derivation of the effective BdG
equation at the TI surface. Here we refine the reasoning of
Ref.~\onlinecite{sau_robustness,sau_preprint}.

The system in question consists of a superconducting slab, described by the
BCS Hamiltonian
$H_{\rm SC}$,
and a topological insulator with Hamiltonian
$H_{\rm TI}$.
These two are separated by a flat interface. Tunneling across this
interface is described by the tunneling Hamiltonian:
$\widehat {\cal T} + \widehat {\cal T}^\dag$.

As the TI has the bulk gap, its low-energy states are located at the
surface. At low energy these 2D states can be approximately described by
the Weyl-Dirac Hamiltonian whose apex is located at the
${\bf M}$
point at the boundary of the Brillouin zone of the TI.

Due to conservation of the spin and the quasi-momentum parallel to the flat
interface, the tunneling Hamiltonian couples a single-electron state
$\left| \phi^{\rm TI}_{{\bf k} \sigma } \right>$
inside the TI with a normal-metal state
$\left| \chi^{\rm m}_{{\bf M+k}, k_z, \sigma} \right>$
inside the superconductor. Here
$\sigma$
is the spin index,
${\bf k}$
is the momentum's components parallel to the barrier measured from
${\bf M}$,
and 
${k}_z$
is the absolute value of the transverse momentum (since scattering at the
interface couples SC states with $k_z$ and $-k_z$, the state
$\left| \chi^{\rm m}_{{\bf M+k}, k_z, \sigma} \right>$
is a boundary-condition-compatible linear combination of both $k_z$ and
$-k_z$ states). The superscript `m' stands for `metal'. Let us denote the
corresponding matrix element by
${\cal T}({\bf k}, k_z)$.
It is assumed to be spin independent.

Invariance of the tunneling Hamiltonian with respect to the spatial
inversion
${\bf r} \rightarrow -{\bf r}$
implies that the same matrix element couples the states
$\left| \phi^{\rm TI}_{-{\bf k} \sigma } \right>$
and
$\left| \chi^{\rm m}_{{\bf -M-k}, k_z, \sigma} \right>$.
Indeed, the quasi-momentum
${\bf M-k}$
of the TI state is equal to the momentum
${\bf -M-k}$
of the state inside the superconductor, modulo the TI's reciprocal lattice
vector
$2{\bf M}$.
Thus, the anomalous term of
$H_{\rm SC}$,
which mixes
$\left| \chi^{\rm m}_{{\bf -M-k}, k_z, \sigma} \right>$
and
$\left| \chi^{\rm m}_{{\bf M+k}, k_z, \sigma} \right>$,
induces a coupling between
$\left| \phi^{\rm TI}_{{\bf k} \sigma } \right>$
and
$\left| \phi^{\rm TI}_{-{\bf k} \sigma } \right>$
in
$H_{\rm eff}$.

In addition, recall that for a tunneling matrix element
${\cal T}$
between certain electronic states there is a tunneling matrix element
$-{\cal T}$
between corresponding charge-conjugated states. This implies that the operator
$\widehat {\cal T}$
in the Nambu representation
Eq.~(\ref{nambu_spinor})
is proportional to
$\tau_z$,
see Eq.~(\ref{tunneling}).

The single-quasiparticle states are given by the following BdG equations
\begin{eqnarray}
\label{app::psi}
\nonumber
\left(
	H_{\textrm{TI}}-\omega
\right)
\Psi_{\textrm{TI}}
+
{\cal{T}}^\dag
\Psi_{\textrm{SC}}&=& 0, \\
\left(
	H_{\textrm{SC}}-\omega
\right)
\Psi_{\textrm{SC}}
+
{\cal{T}}
\Psi_{\textrm{TI}}&=& 0.
\end{eqnarray}
The four-component spinor
$\Psi_{\textrm{SC}}$
($\Psi_{\textrm{TI}}$)
describes a quasiparticle inside the SC (TI).

Solving the second of the Eqs.~\eqref{app::psi} for the wave-function
$\Psi_{\textrm{SC}}$
and substituting the resultant expression in the first one, we derive the
effective BdG equation on the TI surface
\begin{equation}
\label{app::BdG}
\left[
	H_{\textrm{TI}}
	+
	\Sigma(\mathbf{k},\omega)
	-
	\omega
\right]\Psi_{\textrm{TI}}=0,
\end{equation}
where the self-energy $\Sigma$ on the TI surface reads
\begin{eqnarray}
\label{Sigma}
\Sigma(\mathbf{k},\omega)
=
-\frac{1}{L_z} 
\sum_{k_z}
\left|
	{\cal{T}}(\mathbf{k}, k_z)
\right|^2
\tau_z \: G_{\textrm{SC}}(\mathbf{k}, k_z, \omega) \: \tau_z.
\end{eqnarray}
The Green's function for the superconductor
$G_{\textrm{SC}}=\left(H_{\textrm{SC}}-\omega\right)^{-1}$
can be derived with the help of
Eq.~\eqref{HH}:
\begin{equation}
\label{GG}
G_{\textrm{SC}}(\mathbf{k},k_z,\omega)
=
\left[
	\epsilon(\mathbf{k},k_z)\tau_z
	+
	\Delta\tau_x
	-
	\omega \tau_0
\right]^{-1},
\end{equation}
where
$\epsilon(\mathbf{k},k_z)=\left(k^2+k_z^2\right)/2m-\varepsilon_F$.
Substituting
Eq.~\eqref{GG}
in
Eq.~\eqref{Sigma},
one obtains
\begin{equation}\label{SS}
\Sigma(\mathbf{k},\omega)
=
\int\frac{dk_z}{2\pi}
\frac{
	\Delta\tau_x
	-
	\omega\tau_0
	-
	\epsilon\tau_z
     }
     {
	\epsilon^2+|\Delta|^2-\omega^2
     }
\left|{\cal{T}} ({{\bf k}, k_z})\right|^2.
\end{equation}
Here $\tau_0$ is the unit matrix. The tunneling matrix element
${\cal{T}} ({{\bf k}, k_z})$
is assumed to vary slowly as a function of
$k_z$.
Then, transforming the
$k_z$
integral to an energy one, we derive
\begin{equation}
\label{SSS}
\Sigma(\mathbf{k},\omega)
\approx
\lambda(\mathbf{k})
\frac{
	\Delta\tau_x
	-
	\omega\tau_0
     }
     {
	\sqrt{|\Delta|^2-\omega^2}
     }
-
\delta U \tau_z,
\end{equation}
where
$\lambda(\mathbf{k})
=
(\pi/2)\nu(\varepsilon_F,\mathbf{k})
\left|{\cal{T}} ({{\bf k}, k_z})\right|^2$
characterizes the transparency of the interface. The quantity
$\nu(\varepsilon,\mathbf{k})$
quantifies the DOS at a given energy and parallel momentum in the normal
state of the SC. It is equal to
\begin{equation}
\label{app::nu}
\nu(\varepsilon,\mathbf{k})
=
\int\frac{dk_z}{2\pi}
\delta\left[\varepsilon - \epsilon(\mathbf{k},k_z)\right].
\end{equation}
In what follows we are interested in states close to the Dirac cone
$\mathbf{k}=\mathbf{M}$
and ignore the k-dependence of $\lambda$, assuming that
$\lambda({\bf k}) \approx \lambda({\bf M})$.

The correction to the TI Fermi energy due to tunneling is equal to
\begin{eqnarray}
\delta U
=
\int\frac{dk_z}{2\pi}
\frac{
	\epsilon \left|{\cal{T}} ({{\bf k}, k_z})\right|^2
     }
     {
	\epsilon^2+|\Delta|^2-\omega^2
     }
\approx
\lambda(\mathbf{M})
\int^{E_{\rm h}}_{-E_{\rm e}}
\frac{d \epsilon}{\epsilon}
\\
\nonumber
\approx
\lambda ({\bf M}) \ln \frac{E_{\rm h}}{E_{\rm e}}
=
O(\lambda),
\end{eqnarray}
where
$E_{\rm h}$
($E_{\rm e}$)
is the largest hole (electron) energy relative to the Fermi level, and the
integral over
$1/\epsilon$ is taken using the Cauchy principle value. This term is, in
general, non-zero. It was discarded in
Ref.~\onlinecite{sau_robustness},
presumably, because it can be absorbed into the renormalized Fermi energy.
However, if $\lambda$ depends on the position
${\bf r}$
(which is the case for both our system as well as the system of
Ref.~\onlinecite{sau_robustness}),
then
$\delta U$
becomes spatially inhomogeneous as well. Under such circumstances,
$\delta U ({\bf r})$
cannot be absorbed into the Fermi energy, and has to be treated separately.
It should be possible, however, to compensate
$\delta U$
by an external gate potential.

Finally, using
$H_{\textrm{TI}}$
from
Eq.~\eqref{dirac}
and $\Sigma$ from
Eq.~\eqref{SSS},
we rewrite
Eq.~\eqref{app::BdG}
in the explicit
form~\cite{sau_robustness}
\begin{eqnarray}
\label{app::BdG1}
\left(H_{\textrm{eff}}-\omega\right)\Psi_{\textrm{TI}}=0,
\\
H_{\textrm{eff}}
=
\left[
	i\tilde{v}(\omega)
	(\mathbf{\sigma}\cdot\mathbf{\nabla}_{\bf r})
	-
	\tilde{U}(\omega)
\right] \tau_z
+
\tilde{\Delta}(\omega)\tau_x,
\\
\tilde{v}(\omega)
=
\frac{
	v\sqrt{|\Delta|^2-\omega^2}
     }
     {
	\sqrt{|\Delta|^2-\omega^2}+\lambda
     },
\\
\tilde{U}(\omega)
=
\frac{
	(U + \delta U) \sqrt{|\Delta|^2-\omega^2}
     }
     {
	\sqrt{|\Delta|^2-\omega^2}+\lambda
     },
\\
\tilde{\Delta}(\omega)
=
\frac{
	\Delta\lambda
     }
     {
	\sqrt{|\Delta|^2-\omega^2}+\lambda
     }.
\end{eqnarray}

\section{Bogolyubov-de Gennes equations for the radial motion}
\label{app::radial}
In this Appendix we will exploit the cylindrical symmetry of the problem,
to separate angular and radial variables, and to derive the equations for
the radial part of the wave function. The effective Hamiltonian with a
single vortex is equal to
\begin{eqnarray}
\label{Hef1}
H_{\textrm{eff}}
&=&
i\tilde{v}(r)
\exp{\left(i\theta\sigma_z\right)}
\left(
	\sigma_x\partial_r+\frac{1}{r}\sigma_y\partial_\theta
\right)
\tau_z
\\
&+&
|\tilde{\Delta}(r)|\exp {\left( i\frac{\theta}{2}\tau_z \right)}\tau_x
\exp {\left(\!-i\frac{\theta}{2}\tau_z \right)}-\tilde{U}(r)\tau_z\,.
\nonumber
\end{eqnarray}
The first term here is the Weyl-Dirac Hamiltonian in polar coordinates.
The second term corresponds to the anomalous term in the presence of a
vortex.

Equation~(\ref{Hef1})
suggests that it is useful to define a new spinor $\Psi$
\begin{eqnarray}
\Psi_{\textrm{TI}}
=
\exp{\left[i\theta(\tau_z+\sigma_z)/2\right]}\Psi.
\end{eqnarray}
Accordingly, the effective Hamiltonian is transformed as
\begin{eqnarray}
H'_{\textrm{eff}}
=
\exp{\left[-i\theta(\tau_z+\sigma_z)/2\right]}
H_{\textrm{eff}}
\exp{\left[i\theta(\tau_z+\sigma_z)/2\right]}.
\qquad
\end{eqnarray}
The equation for $\Psi$ now reads
\begin{eqnarray}
\label{BdGf}
\Big\{
	i\tilde{v}\left[
			\sigma_x\partial_r
			+
			\frac{1}{r}\sigma_y\partial_\theta
			+
			\frac{i}{2r}\sigma_y ( \tau_z + \sigma_z)
		 \right]
	\tau_z
\nonumber
\\
	+
	|\tilde{\Delta}|\tau_x
	-
	\tilde{U}\tau_z
	-{\omega}
\Big\}\Psi=0\,.
\end{eqnarray}
We look for a solution of
Eq.~\eqref{BdGf}
in the form
$\Psi=e^{i\mu\theta}\Phi^\mu$, where
$\mu=0, \pm 1, \pm 2,...$
is the angular momentum. The values of $\mu$ are integers (not
half-integers) to ensure single-valuedness of
$\Psi_{\rm TI}$.
It is straightforward to show now that the components of the spinor
$\Phi^\mu=(f_1^\mu,f_2^\mu,f_3^\mu,-f_4^\mu)^T$
satisfy
Eq.~(\ref{f1}).

\end{document}